# DYNAMICAL BEHAVIOR OF A SQUID RING COUPLED TO A QUANTIZED ELECTROMAGNETIC FIELD

R. MIGLIORE AND A. MESSINA

*INFM, MIUR and Dipartimento di Scienze Fisiche ed Astronomiche, Via Archirafi 36, I-90123 Palermo Italy*
*E-mail: rosanna@fisica.unipa.it*

In this paper we investigate the dynamical behavior of a SQUID ring coupled to a quantized single-mode electromagnetic field. We have calculated the eigenstates of the combined fully quantum mechanical SQUID-field system. Interesting phenomena occur when the energy difference between the usual symmetric and anti-symmetric SQUID states equals the field energy $\hbar w_F$. We find the low-energy lying entangled stationary states of the system and demonstrate that its dynamics is dominated by coherent Rabi oscillations.

## 1  Introduction

The interest in "macroscopic" quantum effects in circuits involving small Josephson junctions has persisted for many years. One of the motivation was to test whether the law of quantum mechanics, familiar in the microscopic world, apply in macroscopic systems. It is interesting then to try to observe quantum effects involving dynamical variables by which it is possible to describe macroscopically distinct states [1]. The relevant macroscopic degree of freedom is the phase difference between the superconducting order parameters across a junction or the magnetic flux $f$ threading a SQUID ring [2]. Several quantum phenomena such as macroscopic quantum tunneling (MQT) and the resonant tunneling between quantized energy levels in the adjacent wells of the Josephson potential, have been widely studied and experimentally observed, agreeing with theory [3-5]. Another motivation is the strong connection to the theory of quantum information. Circuits involving Josephson junction, in fact, behave as macroscopic two-level systems which can be externally controlled and then may serve as quantum bits (qubits) in quantum information devices [6]. But in order to perform quantum logic operations, it must be possible to prepare and observe quantum superpositions of macroscopically distinct states, coherent oscillations and entangled states of two or several qubits. Superpositions of different flux states have been observed [7-8] and new efforts are made to observe the coherent oscillations between degenerate states [8-10]. Entangled states of several coupled qubits have been created and manipulated in systems involving atoms and high-Q cavity [11]. More recently there have been theoretical studies of entangled states involving Josephson devices. Buisson and Hekking have studied one of the simplest Josephson circuits (a charge qubit coupled to a superconducting resonator) in which entangled states can be realized [12-13] and similar studies concern the states of a SQUID coupled to a resonant cavity [14-16].

In this article, we study a fully quantum-mechanical model for the coupling between a superconducting quantum interference device (a rf-SQUID) and a single mode quantized electromagnetic field. After describing the quantum system we give its Hamiltonian in section 2. In the next section we deduce the low-energy lying entangled stationary states which allows to deduce the time evolution of the system from initial conditions appropriate to this experimental situation. The last section contains a concluding discussion concerning the dynamics of the quantum system and some brief remarks on the possibility of observing these phenomena.

## 2  Rf-SQUID coupled to nonclassical electromagnetic field

We consider a rf-squid exposed to a single mode quantized electromagnetic field. As in ref. [17], the system is assumed to be describable by the following Hamiltonian

$$H = H_{SQUID} + H_F + H_{INT} , \qquad (1)$$

where $H_{SQUID}$ describes the Josephson device, $H_F$ the quantized electromagnetic field and $H_{INT}$ is the coupling term. Below we describe the two subsystems and discuss their contribution to the total Hamiltonian in some details.

A rf- SQUID is a superconducting loop of self-inductance $L$ interrupted by a Josephson junction with capacitance $C$ and critical current $I_C$. An externally applied dc flux $f_x$ biases the system. The dynamics of the SQUID, described in terms of the magnetic flux $f$ threading the ring, are analogous to those of a particle of kinetic energies $Q^2/2C$ subjected to the potential

$$U(f) = \frac{(f - f_X)^2}{2L} - \frac{I_C f_0}{2p} \cos\left(\frac{2pf}{f_0}\right) \qquad (2)$$

where the charge $Q = -i\hbar \partial/\partial f$ on the leads is canonically conjugate to $f$ and $f_0 = h/2e$ is the flux quantum [2]. When the parameter $b_L \equiv 2pLI_C/f_0$ is larger than 1 and $f_x = f_0/2$ the potential $U(f)$ is a symmetric double well with the left and right wells corresponding to the two different senses of rotation of the supercurrent around the loop. Any change in $f_x$ then tilts the potential, resulting in an energy difference $\hbar e$ between the two potential minima (figure 1).

If the barrier height $V_b$ is large compared to $\hbar w_0$, where $w_0$ represent the classical oscillation frequency around the minimum in each well, tunneling does not mix the two lowest flux states with the excited states in the wells. Thus, at very low temperature ($k_B T \ll V_b$), the SQUID behaves as an effective two state system describable in terms of the following reduced Hamiltonian [18,19]:

$$H_S = -\frac{1}{2}\hbar\Delta s_x + \frac{1}{2}\hbar e s_z = \frac{\hbar}{2}\begin{pmatrix} e & \Delta \\ \Delta & -e \end{pmatrix}. \qquad (3)$$

Here $s_x$ and $s_z$ are the Pauli matrices, the off diagonal term $\Delta$ describes the tunneling amplitude between the wells and the basis is formed by the localized states $|R\rangle$ and $|L\rangle$ which are eigenstates of $s_z$ with eigenvalues +1 and –1, respectively. The position operator is $f = (f_0/2)s_z$ and the eigenvalues $\pm f_0/2$ are the position of the localized states. This matrix can be diagonalized. The eigenvalues are $E_\mp = \mp \hbar\sqrt{e^2 + \Delta^2}/2$ and the corresponding normalized eigenvectors are:

$$|-\rangle = c_-|R\rangle + c_- \frac{(e + \sqrt{e^2 + \Delta^2})}{\Delta}|L\rangle \qquad (4)$$

$$|+\rangle = c_+|R\rangle + c_+ \frac{(e - \sqrt{e^2 + \Delta^2})}{\Delta}|L\rangle \tag{5}$$

where $c_{\mp} = \left[1 + \frac{e \pm \sqrt{e^2 + \Delta^2}}{\Delta}\right]^{-1/2}$.

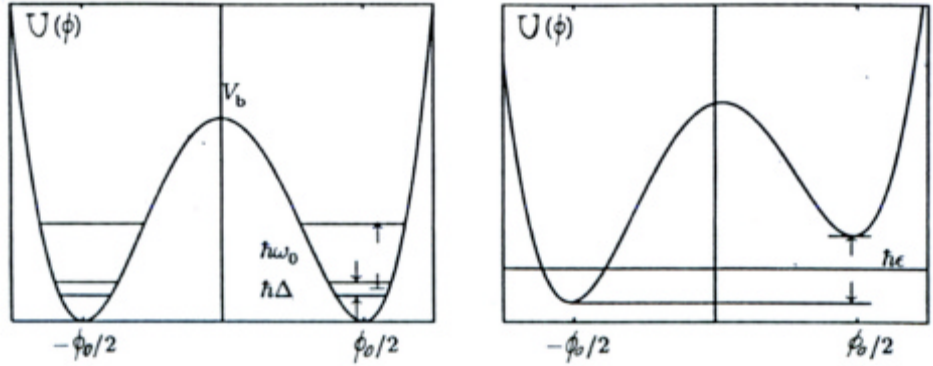

**Figure 1.** Symmetric double well potential (left) with barrier height $V_b$. The separation between the first excited state and the ground state in each well is $\hbar w_0$, and the tunnel splitting is $\hbar\Delta$. The biased double well with detuning energy $\hbar e$ is sketched on the right

The quantized electromagnetic field can be described in terms of a harmonic oscillator with characteristic frequency $w_F$. Then the relative Hamiltonian assumes the standard form

$$H_F = \hbar w_F\left(a^\dagger a + 1/2\right), \tag{6}$$

where the photon annihilation and creation operators $a$ and $a^\dagger$ define the conjugate field operators $Q_F$ and $f_F$ as:

$$Q_F = -i\sqrt{\hbar w_F/2}\left(a - a^\dagger\right) \tag{7}$$

$$f_F = \sqrt{\hbar/2w_F}\left(a + a^\dagger\right). \tag{8}$$

Finally, the inductive coupling between the electromagnetic field and the rf-SQUID is simply described by the flux-flux interaction term

$$H_{INT} = \frac{2k}{L} f f_F \tag{9}$$

where $k$ is an adimensional coupling parameter, typically of the order of 0.01.

The Hamiltonian for the fully quantized rf SQUID-electromagnetic mode system can therefore be written down as follows:

$$H = -\frac{1}{2}\hbar\Delta s_x + \frac{1}{2}\hbar e s_z + \hbar w_F(a^\dagger a + \frac{1}{2}) + \frac{2k}{L}\frac{f_0}{2}\sqrt{\frac{\hbar}{2w_F}}(a + a^\dagger)s_z. \tag{10}$$

To consider the dynamics of our system taking place essentially in the low-lying energy states subspace appears experimental meaningful in the context of the problem under scrutiny. For this reason we drastically simplify our problem reducing $H$ to a finite dimensional matrix form.

## 3  The reduced Hamiltonian: eigenstates and entanglement

As previously remarked, if $f_x \approx f_0/2$ we are legitimated to study the dynamics of the rf-SQUID in the two dimensional subspace generated by its anti-symmetric and symmetric states $|-\rangle$ and $|+\rangle$ respectively.

Moreover we assume that $\hbar w_F$ is of the order of the tunnel splitting $\hbar\Delta$ so that in the weak coupling limit we may represent the total Hamiltonian $H$ in the reduced Hilbert space spanned by the four states $|-,0\rangle$, $|-,1\rangle$, $|+,0\rangle$ and $|+,1\rangle$, $|0\rangle$ and $|1\rangle$ being the ground and the first excited states of the electromagnetic field respectively. Thus the operator $H$ in equation (10) is substituted by the following hermitian matrix

$$H = \begin{pmatrix} E_0 & -Ae & 0 & A\Delta \\ -Ae & E_1 & A\Delta & 0 \\ 0 & A\Delta & E_2 & Ae \\ A\Delta & 0 & Ae & E_3 \end{pmatrix} \quad (11)$$

where $E_0 = E_- + \hbar w_F/2$, $E_1 = E_- + 3\hbar w_F/2$, $E_2 = E_+ + \hbar w_F/2$, $E_3 = E_+ + 3\hbar w_F/2$ and $A = \frac{k}{L}\left(\sqrt{\frac{\hbar}{2w_F}}\right)\frac{f_0}{\sqrt{e^2+\Delta^2}}$. If now we consider an externally applied flux $f_x = f_0/2$ and a field frequency $w_F = \Delta$, we have $E_0 = e = 0$, $E_1 = E_2 = \hbar w_F$ and $E_3 = 2\hbar w_F$, i.e. the states $|-,1\rangle$ and $|+,0\rangle$ have the same energy. In such resonant conditions the Hamiltonian (11) can be cast in the form

$$H_d = \begin{pmatrix} 0 & 0 & 0 & B\Delta \\ 0 & \hbar w_F & B\Delta & 0 \\ 0 & B\Delta & \hbar w_F & 0 \\ B\Delta & 0 & 0 & 2\hbar w_F \end{pmatrix} \quad (12)$$

where $B = \frac{k}{L}\sqrt{\frac{\hbar}{2w_F}}\frac{f_0}{w_F}$. It is interesting to analyze the central 2×2 matrix block which describes the entanglement between the states $|-,1\rangle$ and $|+,0\rangle$. The eigenstates of this block are

$$|u_1\rangle = [|+,0\rangle - |-,1\rangle]/\sqrt{2} \quad (13)$$
$$|u_2\rangle = [|+,0\rangle + |-,1\rangle]/\sqrt{2} \quad (14)$$

with eigenvalues $l_1 = (\hbar - B)w_F$ and $l_2 = (\hbar + B)w_F$. This eigenstates are example of maximally entangled states of the total system. It is immediate to express the old basis $|-,1\rangle$ and $|+,0\rangle$ in terms of the new one as follows:

$$|+,0\rangle = [|u_1\rangle + |u_2\rangle]/\sqrt{2} \tag{15}$$

$$|-,1\rangle = -[|u_1\rangle - |u_2\rangle]/\sqrt{2} \tag{16}$$

It is now very simple to written down the time evolution of the combined system, initially prepared in the state $|y(t=0)\rangle = |+,0\rangle$:

$$|y(t)\rangle = \frac{1}{\sqrt{2}}[|u_1\rangle \exp(-il_1 t/\hbar) + |u_2\rangle \exp(-il_{21} t/\hbar)] \tag{17}$$

This expression clearly evidence the existence of coherent Rabi oscillations between the states $|-,1\rangle$ and $|+,0\rangle$ corresponding to the absorption and emission of a quantum of energy by the SQUID. Such a periodic behaviour, dominating the dynamics of the system in the low lying energy subspace, is a direct consequence of the entanglement get established between the matter and radiation subsystems in the stationary states $|u_1\rangle$ and $|u_2\rangle$. The probability

$$P(t) = |\langle 1,-|y(t)\rangle|^2 = \frac{1}{2}\left[1 - \cos\left(2\frac{Bw_F}{\hbar}t\right)\right] \tag{18}$$

of finding the system, initially prepared in the state $|+,0\rangle$, in the state $|-,1\rangle$ after a time $t$ is showed in figure 2. The Rabi frequency is $\Omega = 2Bw_F/\hbar$ and its numerical estimation is reported in the next section.

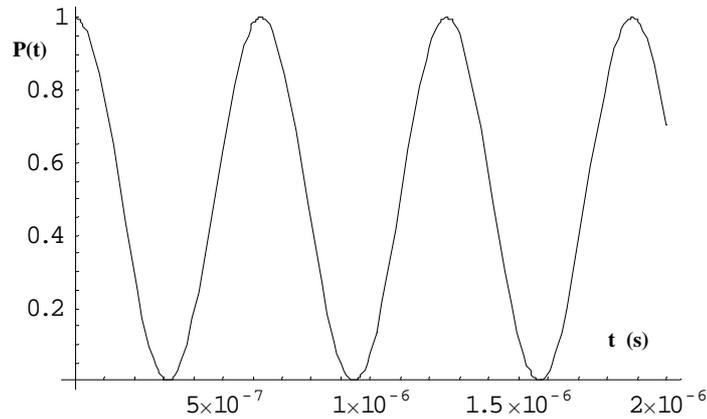

**Figure 2.** Probability P(t) to find the system, prepared at t=0 in the state $|+,0\rangle$, in the state $|-,1\rangle$ after a time t.

## 4  Discussion

In this paper we have considered a symmetric double well wherein the localized states in the wells $|R\rangle$ and $|L\rangle$ have the same energy. Moreover we assume that the field frequency $w_F$ is equal to the off diagonal term $\Delta$ describing the tunneling frequency between the wells. In order to estimate its value we consider that realistic experimental values [8, 19] of $C$ and $L$ for a SQUID working at $T = 10\ mK$ are $10^{-13} \div 10^{-14}\ F$ and $20 \div 100\ pH$ respectively. With these values the characteristic time of oscillation between localized states in the wells is $2p\Delta^{-1} \approx 30\ ns$. This means that choosing $w_F \approx 2 \cdot 10^8\ rad\ s^{-1}$, the frequency $\Omega = 2Bw_F/\hbar$ characterizing the Rabi oscillations between the two state $|+, 0\rangle$ and $|-, 1\rangle$ corresponds to periods $t_R$ of the order of $10^{-7} s$. We conclude emphasizing that the observability of coherent Rabi oscillations in such systems seems to be in the grasp of the experimentalists since the decoherence time related to the electromagnetic mode decay is longer than $t_R$ in view of the current Q factors of the electromagnetic resonators at the indicated frequency.